\documentclass[prl,aps,superscriptaddress,preprint,floatfix,
nofootinbib]{revtex4}

\usepackage{amsmath,amsfonts,amssymb}
\usepackage{graphicx}

\setcitestyle{round}
\def\3{2.8in}    %used for figure widths
\def\2{2.5in}
\def\4{3.0in}

\def \beq {\begin{equation}}
\def \eeq {\end{equation}}
\begin{document}

\title{Topological Quantum Phase Transition and 3D Texture Inversion in a Tunable Topological Insulator}
\author{Su-Yang Xu}\affiliation {Joseph Henry Laboratory of Physics, Department of Physics, Princeton University, Princeton, New Jersey 08544, USA}
\author{Y. Xia}\affiliation {Joseph Henry Laboratory of Physics, Department of Physics,
  Princeton University, Princeton, New Jersey 08544, USA}
\author{L. A. Wray}\affiliation{Joseph Henry Laboratory of Physics, Department of Physics,
  Princeton University, Princeton, New Jersey 08544, USA}\affiliation {Advanced Light Source, Lawrence Berkeley National Laboratory, Berkeley, California 94305, USA}
\author{D. Qian}\affiliation {Joseph Henry Laboratory of Physics, Department of Physics, Princeton University, Princeton, New Jersey 08544, USA}
\author{S. Jia}\affiliation {Department of Chemistry, Princeton University, Princeton, New Jersey 08544, USA}
\author{J. H. Dil}\affiliation {Swiss Light Source, Paul Scherrer Institute, CH-5232, Villigen, Switzerland}\affiliation {Physik-Institute, Universitat Zurich-Irchel, CH-8057 Zurich, Switzerland}
\author{F. Meier}\affiliation {Swiss Light Source, Paul Scherrer Institute, CH-5232, Villigen, Switzerland}\affiliation {Physik-Institute, Universitat Zurich-Irchel, CH-8057 Zurich, Switzerland}
\author{J. Osterwalder}\affiliation {Physik-Institute, Universitat Zurich-Irchel, CH-8057 Zurich, Switzerland}
\author{B. Slomski}\affiliation {Swiss Light Source, Paul Scherrer Institute, CH-5232, Villigen, Switzerland}\affiliation {Physik-Institute, Universitat Zurich-Irchel, CH-8057 Zurich, Switzerland}
\author{H. Lin}\affiliation {Department of Physics, Northeastern University, Boston, Massachusetts 02115, USA}
\author{R. J. Cava}\affiliation {Department of Chemistry, Princeton University, Princeton, New Jersey 08544, USA}
\author{M. Z. Hasan}\affiliation {Joseph Henry Laboratory of Physics, Department of Physics,
  Princeton University, Princeton, New Jersey 08544, USA} \affiliation {Advanced Light Source, Lawrence Berkeley National Laboratory, Berkeley, California 94305, USA} \affiliation {Princeton Institute for Science and Technology of Materials, Princeton School of Engineering and Applied Science, Princeton University,  Princeton University, Princeton, New Jersey 08544, USA}\affiliation {Princeton Center for Complex Materials, Princeton University, Princeton, New Jersey 08544, USA}

%76.20.+q,
%85.35.Gv,
%82.20.Xr,
%76.60.Jx,
%39.20.+q}

\begin{abstract}

\textbf{The recently discovered three dimensional or bulk topological insulators are expected to exhibit exotic
quantum phenomena. It is believed that a trivial insulator can be twisted into a topological state by modulating the
spin-orbit interaction or the crystal lattice via odd number of band inversions, driving the system through a topological quantum phase transition.
By directly measuring the topological invariants (for method to directly measure {$\nu_0$} of Fu-Kane, see Hsieh \textit{et.al.,} Science 323, 919 (2009) at http://www.sciencemag.org/content/323/5916/919.abstract) we report the observation of a topological phase transition in a tunable spin-orbit system BiTl(S$_{1-\delta}$Se$_{\delta}$)$_2$ (which is an analog of the most studied topological insulator Bi$_2$Se$_3$, see Xia \textit{et.al.,} Nature Phys. 5, 398 (2009) at http://www.nature.com/nphys/journal/v5/n6/full/nphys1294.html and spin-momentum locked topological-order at Hsieh \textit{et.al.,} Nature 460, 1101 (2009)) where the topological insulator formation process is visualized for the first time. In the topological state, vortex-like polarization states are observed to exhibit 3D vectorial textures, which collectively feature a chirality transition of its topological spin-textures as the spin-momentum locked (topologically ordered) electrons on the surface go through the zero carrier density point. Such phase transition and texture chirality inversion can be the physical basis for observing fractional charge ($\pm$e/2) and other related fractional topological phenomena.}

\end{abstract}

\maketitle

Topological insulators (TIs) in three dimensions are nonmagnetic insulators with novel surface states (SSs) that are a consequence of the nontrivial topology of electronic wavefunctions in the bulk of the materials (\textit{1-5}). Their experimental discoveries first in Bi-Sb semiconductors (Hsieh \textit{et.al.,} 2007, 2008), then in Bi$_2$Se$_3$ (Xia \textit{et.al.,} 2008, 2009) and related materials have led to the exploration of topological quantum phenomena at modest temperatures and without the requirement of any applied magnetic field (\textit{6-16}). The potential for the realization of fractional quasi-particles and exploitation of the role of spin-momentum locking leading to quantum phases (such as topological non-Abelian phases) in charge dynamics are some of the intriguing current theoretical proposals (\textit{17-24}).

Strong TIs are distinguished from ordinary insulators by a finite topological quantum number or an invariant ($\nu_0$) or equivalently an axion angle parameter ($\theta$). The value of $\nu_0$ or $\theta$ depends on the parity eigenvalues of the wavefunctions in time reversal symmetric materials (\textit{1, 2}), and can be determined from spin texture details of the spin-orbit surface states that form when the bulk is terminated as demonstrated by Hsieh \textit{et al}. (\textit{6, 7}). In particular, a $\nu_0=1=\theta/\pi$ topology requires the terminated surface to have a Fermi surface (FS) that supports a nonzero Berry's phase (an odd as opposed to an even multiple of $\pi=\theta=$axion angle), which is not realizable in an ordinary spin-orbit or trivial ($\nu_0=0=\theta$) insulator. It is believed that a trivial insulator can be twisted into a $\nu_0=1=\theta/\pi$ topological state by signaling the appearance of unusual spin vortex-like [or Skyrmion-like (\textit{1})] arrangements via increasing spin-orbit interaction or by modulating the lattice parameters, thereby driving the system through a topological quantum phase transition (\textit{1, 2}). However, the TIs Bi$_2$X$_3$ (X=Se,Te) cannot be tuned out from a trivial insulator version without a structural phase transition; in the Bi-Sb semiconductors, the topological phase transition is masked by an intervening band [the ``H'' band (\textit{1})] and the lack of gating control necessary for its observation.

We demonstrate the existence of a topological phase transition in TlBi(S$_{1-\delta}$Se$_{\delta}$)$_2$ which, as we show, is a fully tunable topological analog of Bi$_2$Se$_3$. By effectively varying the sulfur to selenium ratio ($\delta$), both the spin-orbit strength and lattice parameters are effectively tuned (\textit{25}). Because the topological order in TlBi(S$_{1-\delta}$Se$_{\delta}$)$_2$ originates from Bi and Se atoms and to illustrate an analogy with the known topological insulator Bi$_2$Se$_3$ (\textit{6, 8, 11}) we rewrite the formulae as BiTl(S$_{1-\delta}$Se$_{\delta}$)$_2$. Figure 1A presents systematic photoemission measurements of electronic states that lie between a pair of time-reversal invariant points or Kramers' points ($\bar{\Gamma}$ and $\bar{M}$) obtained for a series of compositions of the spin-orbit material BiTl(S$_{1-\delta}$Se$_{\delta}$)$_2$. As the selenium concentration is increased, the low-lying bands separated by a gap of energy 0.15eV at $\delta=0.0$ are observed to approach each other and the gap decreases to less than 0.05eV at $\delta=0.4$. Both bands demarcating the gap show three dimensional dispersion where binding energies vary with momentum perpendicular to the surface, $k_z$, [as probed via varying incident photon energy (\textit{25})] and roughly correspond to the expected position for the valence and conduction bands. The absence of SSs within the bulk gap suggests that the compound is topologically trivial ($\nu_0=0=\theta$) for composition range of $\delta=0.0$ and $\delta=0.4$. Starting from $\delta=0.6$, a linearly dispersive band connecting the bulk conduction and valence bands emerges which threads across the bulk band gap. Incident photon energy modulation studies support the assignment of these Dirac-like bands to be of surface origin (\textit{25}). Moreover, while the band continua in the composition range of $\delta=0.0$ to $\delta=0.4$ are degenerate, the Dirac-like bands at $\delta=0.6$ and beyond are spin polarized (see Figs. 2, 3, and 4). The system enters a topologically non-trivial phase upon the occurrence of an electronic phase transition between $\delta=0.4$ and $\delta=0.6$ at temperatures below 15K. While the system approaches the transition from the conventional or no-surface-state side ($\delta=0.4$), both energy dispersion and FS mapping (Fig. 1, A and $\delta=0.4$) show that the spectral weight at the outer boundary of the bulk conduction band continuum which corresponds to the loci where the Dirac SSs would eventually develop becomes much more intense; however, that the surface remains gapped at $\delta=0.4$ suggests that the material is still on the trivial side. Finer control of bulk compositional variation does not allow locating a precise value for the transition; this could also be a consequence of an intrinsically broader topological transition. A critical signature of a topological phase transition is that the material turns into an indirect bulk band gap material as conjectured previously (\textit{1}). As $\delta$ varies from 0.0 to 1.0 (Fig. 1C), the dispersion of the valence band evolves from a ``$\Lambda$''-shape to an ``M''-shape with a ``dip'' at the $\bar{\Gamma}$ point ($k=0$); the $\delta=0.0$ compound features a direct band gap in its bulk, whereas the $\delta=1.0$ indicates a slightly indirect gap. The overall experimental evolution of the spin-orbit groundstate is presented in Fig. 2A.

Now we systematically study the end product of the transition - BiTl(S$_{0}$Se$_{1}$)$_2$ and explore its spin and polarization properties far away from the Dirac node where non-linear spin-orbit terms are also important. Such terms correspond to analogs of cubic and higher order Dresselhaus effects arising from the symmetry of the crystal potential. Without these effects the SSs form a cone which is isotropic in momentum space ($k_x$, $k_y$) and hence generate a circular Fermi contour (\textit{2}). Spins tangentially arranged on such a circular FS lead to a Berry's phase of $\pi$, which is also a measure of the topological invariant and the axion angle ($\nu_0=1=\theta/\pi$) of the system (\textit{1, 4, 6}). However, the details of the symmetry and bulk crystal potential can significantly deform the SSs from that of a circle. We explored the Fermi contour of BiTl(S$_{0}$Se$_{1}$)$_2$ over a large energy range. This sample features an almost perfectly hexagonal FS at its native (as-grown) Fermi level (binding energy $E_B=0.01$eV), whereas the constant energy contour reverts to isotropic circular shapes when approaching the Dirac node ($E_B=0.25$eV and $E_B=0.50$eV). A more extreme example is the TI, n-type bulk doped Bi$_2$Te$_3$, which features a highly warped concave-in snowflake-shaped FS because of its large $k_F$ and a small band gap (\textit{25}). In order to understand the relationship between these Dirac cone deformations and the topological invariants and axion angles ($\nu_0$ and $\theta$), we utilized spin-resolved photoemission spectroscopy (spin-ARPES) (\textit{26, 27}). Such spin-resolved study is also critical as Hall transport data are now possible that rely heavily on surface state topology and spin configuration data for interpretation (\textit{15}).

We have carried out spin polarization texture measurements over a large binding energy scale to capture the non-linear regime that can be accessed by gating for transport measurements. For simplicity, we show the results of the hexagonal FS above the Dirac node and one of the circular FSs below the Dirac node of BiTl(S$_{0}$Se$_{1}$)$_2$ (see Fig. 2,
B and E). Figure 2F shows the measured out-of-plane spin-polarization of cuts C and E (binding energy and momentum
direction are defined in Fig. 2, D and E, respectively). No significant out-of-plane spin-polarization component is
observed within the experimental resolution for cuts as these. The in-plane measurements, on the other hand, show large
polarization amplitudes (Fig. 3A), suggesting that the spin texture is mostly two dimensional. Based on the data, we determine the direction of full 3D spin vectors following a two-step routine (\textit{25, 28}). Because $P_z=0$ (Fig. 2F), the out-of-plane polar angles are all close to $90^{\circ}$, so only in-plane azimuthal angles are shown in Fig. 3B. On the hexagonal Fermi contour which is located above the Dirac node (Fig. 3B), the spin vectors obtained from the polarization measurements show that the groundstate features a 2D inplane left-handed chirality spin vortex (a Skyrmion in momentum space, Fig. 3, B and C). The direction of the spin is roughly perpendicular to the momentum space track that connects the $\bar{\Gamma}$ point and the momentum point location of the spin on the FS, rather than being tangential to the Fermi contour as it is expected for an ideal Dirac cone. Our data also show that the spin texture below the Dirac node is also vortex-like, but features right-handed chirality. Therefore, when the system is chemically tuned through the zero carrier density (Dirac node), the chirality of spin vortex gets inverted as seen in the data. A systematic method of surface chemical potential tuning has been demonstrated in (\textit{6}) which is also applicable here (\textit{25}). However, it is not physically possible to realize this chirality inversion or chemical tuning in Bi$_2$Te$_3$, because the node is buried under other bands.

The observed chirality inversion of the surface spin texture indicates a $180^{\circ}$ turn-around of the spin-momentum locking profile in moving chemical potential across the Dirac node (Fig. 4, A and B). Above the Dirac node, a quasi-particle moving in $+k(+\hat{x})$ direction is locked with $+\hat{y}$ spin-polarization state, whereas below the node, the $+k$ moving quasi-particle state is locked with $-\hat{y}$ spin. These spin-polarization states locked to specific momentum states open up many new possibilities for electrical manipulation of spin in a topological device. Manipulation of Fermi level, e.g., through electrical gating, band-structure engineering or chemical doping-induced gating (\textit{6, 14, 29}) of a TI would allow one to directly observe the consequences of chirality inversion in the quasi-particle dynamics and interference (Fig. 3). One particularly interesting proposal involves electrical or chemical gating of both top and bottom surfaces of a TI thin film, resulting in a left-handed chirality (LHC) electron-like FS on the top and a right-handed chirality (RHC) hole-like FS of Dirac SSs on the bottom. The interaction between LHC and RHC SSs will lead to excitons at low temperatures with topological properties; the detection of such excitons would allow measuring the topologically protected fractionalized charge of $\pm$e/2, which can be implemented by optical methods on a device that incorporates the texture inversion (\textit{17}) possible within this materials class (Fig. 4). To show how these spin textures are relevant in understanding and interpreting the quasi-particle transport such as the anomalous Hall experiments (\textit{15}) involving the TI surfaces, we explore the key scattering processes on the hexagonal and circular surface FSs observed in our data. Scattering profiles (Fig. 3D), estimated based on the measured spin-ARPES FSs and the topological spin textures of BiTl(S$_{0}$Se$_{1}$)$_2$, yield the probability of an electron being scattered in momentum transfer (scattering vector) \textbf{q} space (\textit{25, 31-32}). Scattering vector \textbf{q} is defined as the momentum transfer from one point on a FS ($|\vec{k_F}_1>$) to another $|\vec{k_F}_2>$, thus, \textbf{q}$=|\vec{k_F}_1>-|\vec{k_F}_2>$. We consider the spin texture at a particular binding energy and estimate the spin independent and dependent scattering profile (Fig. 3D). The resulting spin dependent scattering shows a suppression of elastic backscattering on the BiTl(S$_{0}$Se$_{1}$)$_2$ surface consistent with its topological order (\textit{25}). Interestingly, for the hexagonal case, a six-fold symmetric quasi-particle scattering profile is seen, indicating the possibility for a fluctuating spin-density-wave type magnetic instability which connects the parallel pieces of the observed spin-polarized FS (Fig. 3, C and D). This surface magnetic instability is expected to contribute strong fluctuations in the spin transport channel. Based on a mean field Zener theory, it has recently been shown that a helical magnetic order (HMO) is expected if the surface state topology is in the 3D vectorial spin texture (Fig. 4D) regime, the system can then achieve unusually high critical temperature, even up to $T_{HMO}\sim100K$ (\textit{30}). Topological quantization in such a system might then survive at a temperature several orders of magnitude higher than the conventional quantum Hall systems.
\vspace{2cm}

\begin{flushleft}
\textbf{References and Notes}\\
\end{flushleft}
1. M. Z. Hasan, C. L. Kane, Topological insulators. \textit{Rev. Mod. Phys.} $\mathbf{82}$, 3045 (2010).\\
2. L. Fu, C. L. Kane, E. J. Mele, Topological insulators in three dimensions. \textit{Phys. Rev. Lett.} $\mathbf{98}$, 106803 (2007).\\
3. J. E. Moore, Topological insulators: the next generation. \textit{Nature Physics} $\mathbf{5}$, 378 (2009).http://www.nature.com/nphys/journal/v5/n6/full/nphys1294.html \\
4. X.-L. Qi, T. L. Hughes, S.-C. Zhang, Topological field theory of time-reversal invariant insulators. \textit{Phys. Rev. B} $\mathbf{78}$, 195424 (2008).\\
5. M. Z. Hasan, H. Lin, A. Bansil, Warping the cone on a topological insulator. \textit{Physics} $\mathbf{2}$, 108 (2009) and http://www.nature.com/nphys/journal/v5/n6/full/nphys1294.html \\
6. D. Hsieh \textit{et al}., A tunable topological insulator in the spin helical Dirac transport regime. \textit{Nature} $\mathbf{460}$, 1101 (2009); and D. Hsieh \textit{et al}., A topological Dirac insulator in a quantum spin Hall phase. \textit{Nature} $\mathbf{452}$, 970 (2008). Work completed and submitted in 2007, also see,
http://online.itp.ucsb.edu/online/motterials07/hasan/ (2007)
\\
7. D. Hsieh \textit{et al}., Observation of unconventional quantum spin textures in topological insulators. \textit{Science} $\mathbf{323}$, 919 (2009).\\
8. Y. Xia \textit{et al}., Observation of a large-gap topological-insulator class with a single Dirac cone on the surface. \textit{Nature Physics} $\mathbf{5}$, 398 (2009). Y. Xia \textit{et al}., Electrons on the surface of Bi$_2$Se$_3$ form a topologically-ordered two dimensional gas with a non-trivial Berry's phase. http://arxiv.org/abs/0812.2078 (2008). also see,
http://online.itp.ucsb.edu/online/motterials07/hasan/ (2007) \\
9. Y. L. Chen \textit{et al}., Massive Dirac fermion on the surface of a magnetically doped topological insulator. \textit{Science} $\mathbf{329}$, 659 (2010).\\
10. S. V. Eremeev, Y. M. Koroteev, E. V. Chulkov, Ternary thallium-based semimetal chalcogenides Tl-V-VI2 as a new class of 3D topological insulators. \textit{JETP Lett.} $\mathbf{91}$, 594 (2010).\\
11. H. Lin \textit{et al}., Single-Dirac-Cone topological surface states in the TlBiSe$_2$ class of topological semiconductors. \textit{Phys. Rev. Lett.} $\mathbf{105}$, 036404 (2010).\\
12. T. Sato \textit{et al}., Direct evidence for the Dirac-cone topological surface states in the ternary Chalcogenide TlBiSe$_2$. \textit{Phys. Rev. Lett.} $\mathbf{105}$, 136802 (2010).\\
13. Y. Chen \textit{et al}., Single Dirac cone topological surface state and thermoelectric property of compounds from a new topological insulator. \textit{Phys. Rev. Lett.} $\mathbf{105}$ 266401 (2010).\\
14. J. Chen \textit{et al}., Gate-voltage control of chemical potential and weak antilocalization in Bi$_2$Se$_3$. \textit{Phys. Rev. Lett.} $\mathbf{105}$, 176602 (2010).\\
15. D.-X. Qu \textit{et al}., Quantum oscillations and Hall anomaly of surface states in a topological insulator. \textit{Science} $\mathbf{329}$, 821 (2010).\\
16. L. A. Wray \textit{et al}., Observation of topological order in a superconducting doped topological insulator. \textit{Nature Physics} $\mathbf{6}$, 855 (2010).\\
17. B. Seradjeh, J. E. Moore, M. Franz, Exciton condensation and charge fractionalization in a topological insulator film. \textit{Phys. Rev. Lett.} $\mathbf{103}$, 066402 (2009).\\
18. N. Bray-Ali, L. Ding, S. Haas, Topological order in paired states of fermions in two dimensions with breaking of parity and time-reversal symmetries. \textit{Phys. Rev. B} $\mathbf{80}$, 180504(R) (2009).\\
19. J. Linder, Y. Tanaka, T. Yokoyama, A. Sudbo, N. Nagaosa, Unconventional superconductivity on a topological
insulator. \textit{Phys. Rev. Lett.} $\mathbf{104}$, 067001 (2010).\\
20. L. A. Wray \textit{et al}., A topological insulator surface under strong Coulomb, magnetic and disorder perturbations. \textit{Nature Physics} $\mathbf{7}$, 32 (2011).\\
21. D. Pesin, L. Balents, Mott physics and band topology in materials with strong spin-orbit interaction. \textit{Nature Physics} $\mathbf{6}$, 376 (2010).\\
22. M. Kargarian, J. Wen, G. A. Fiete, Competing exotic topological insulator phases in transition metal oxides on the Pyrochlore lattice with distortion. http://arxiv.org/abs/1101.0007 (2011).\\
23. W. Witczak-Krempa, T. Choy, Y. Kim, Gauge field fluctuations in three-dimensional topological Mott insulators. \textit{Phys. Rev. B} $\mathbf{82}$, 165122 (2010).\\
24. D. Hsieh \textit{et al}., A topological Dirac insulator in a quantum spin Hall phase. \textit{Nature} $\mathbf{452}$, 970 (2008).\\
25. Materials and methods are available as supporting material on Science Online.\\
26. M. Hoesch \textit{et al}., Spin-polarized Fermi surface mapping. \textit{J. Electron Spectrosc. Relat. Phenom.} $\mathbf{124}$, 263 (2002).\\
27. J. H. Dil, Spin and angle resolved photoemission on non-magnetic low-dimensional systems. \textit{J. Phys. Condens. Matter} $\mathbf{21}$, 403001 (2009).\\
28. Y. Xia \textit{et al}., Electrons on the surface of Bi$_2$Se$_3$ form a topologically-ordered two dimensional gas with a non-trivial Berry's phase. http://arxiv.org/abs/0812.2078 (2008).\\
29. T. Ohta, A. Bostwick, T. Seyller, K. Horn, E. Rotenberg, Controlling the electronic structure of bilayer graphene. \textit{Science} $\mathbf{313}$, 951 (2006).\\
30. J. H. Jiang, S. Wu, Spin susceptibility and helical magnetic order at the edge of topological insulators due to Fermi surface nesting. http://arxiv.org/abs/1012.1299v2 (2010).\\
31. P. Blaha, K. Schwarz, G. Madsen, D. Kvasnicka, J. Luitz, Computer Code WIEN2K. (Vienna University of Technology 2001).\\
32. P. Roushan \textit{et al}., Topological surface states protected from backscattering by chiral spin texture. \textit{Nature} $\mathbf{460}$, 1106 (2009).\\
33. We acknowledge discussions with M. Neupane, A. Kiatev and D. Haldane. The Synchrotron X-ray-based measurements are supported by the Basic Energy Science of US DOE (DE-FG-02-05ER46200, AC03-76SF00098 and DE-FG02-07ER46352). M.Z.H. acknowledges Visiting Scientist support from LBNL. Materials growth and characterization are supported by NSF-DMR-1006492. M.Z.H. acknowledges additional support from the A. P. Sloan Foundation.

\newpage
\begin{figure*}
%\centering
\includegraphics[width=17cm]{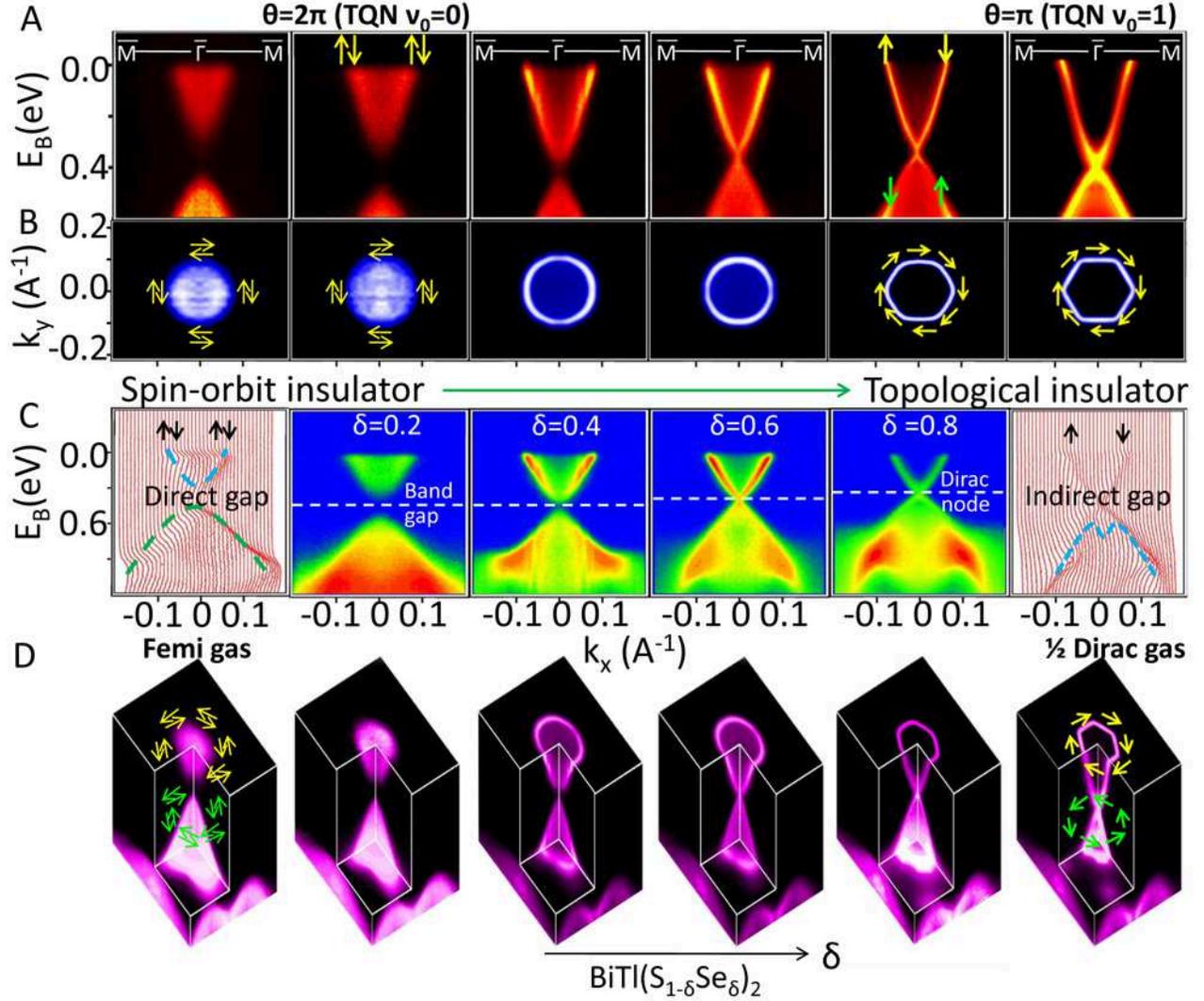}
\caption{\textbf{Topological phase transition.} \textbf{(A)} High resolution ARPES dispersion maps along the $\bar{\Gamma}-\bar{M}$ momentum space line starting from a spin-orbit band insulator (left-most panel) to a topological insulator (right-most panel). Band insulators (non-inverted) and topological insulators (inverted) are characterized by $\theta=2\pi$ and $\pi$ ( $\nu_0$ =1), respectively. \textbf{(B)} ARPES mapped native FSs for different chemical compositions ($\delta=0$ to $1$). \textbf{(C)} Left- and right-most: Energy-distribution curves for stoichiometric compositions $\delta=0$ and $\delta=1$. Middle: ARPES spectra indicating band gap $\delta=0.2$ to $\delta=0.8$. \textbf{(D)} Evolution of electronic ground state (3D band topology) imaged over a wide energy (vertical), spin (arrows) and momentum (horizontal plane) range. Spin textures are indicated by yellow (green) arrows above (below) the Dirac node. Each arrow represents the net polarization direction on a k-space point on the corresponding FS.}
\end{figure*}

\newpage
\begin{figure*}
\centering
\includegraphics[width=17cm]{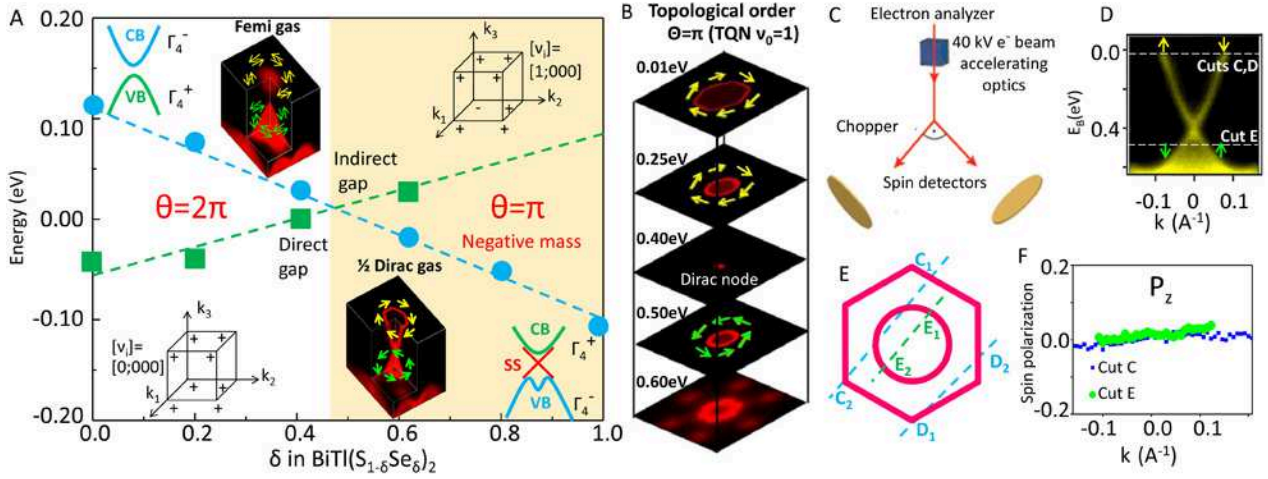}
\caption{\textbf{Evolution of spin-orbit groundstate and spin texture.} \textbf{(A)} Energy levels of $\Gamma_4^-$ (blue circles) and $\Gamma_4^+$ (green circles) bands are obtained from ARPES measurements as a function of composition $\delta$. CB: conduction band; VB: valence band. Parity eigenvalues (+ or -) of Bloch states (\textit{25}) are
shown. The topological invariants, $\nu_i$, obtained from the parity eigenvalues are presented as [$\theta/\pi=\nu_0;\nu_1\nu_2\nu_3$] where $\theta=\pi\nu_0$ is the axion angle and $\nu_0$ is the strong invariant (\textit{1-4}). \textbf{(B)} FS topology evolution of BiTl(S$_{0}$Se$_{1}$)$_2$ across the Dirac node. The corresponding binding energies of constant energy contours are indicated. Observed spin textures are schematically indicated by arrows. \textbf{(C)} Experimental scattering geometry utilized to measure the spin-polarization components presented in \textbf{(B)}. \textbf{(D)} ARPES measured dispersion along the $\bar{\Gamma}-\bar{M}$ momentum space cut. The
binding energies used for the cuts are as follows: $E_B$(cuts C,D)=0.01eV, $E_B$(cut E)=0.50eV. \textbf{(E)} A map of the momentum space cuts C, D and E across the FSs for the spin polarization measurements. The hexagonal (circular) FS is located 0.40eV (0.10eV) above (below) the spin degenerate node. \textbf{(F)} Measured out-of-plane spin-polarization profile of cuts C and E show only weak modulations.}
\end{figure*}

\begin{figure*}
%\centering
\includegraphics[width=17cm]{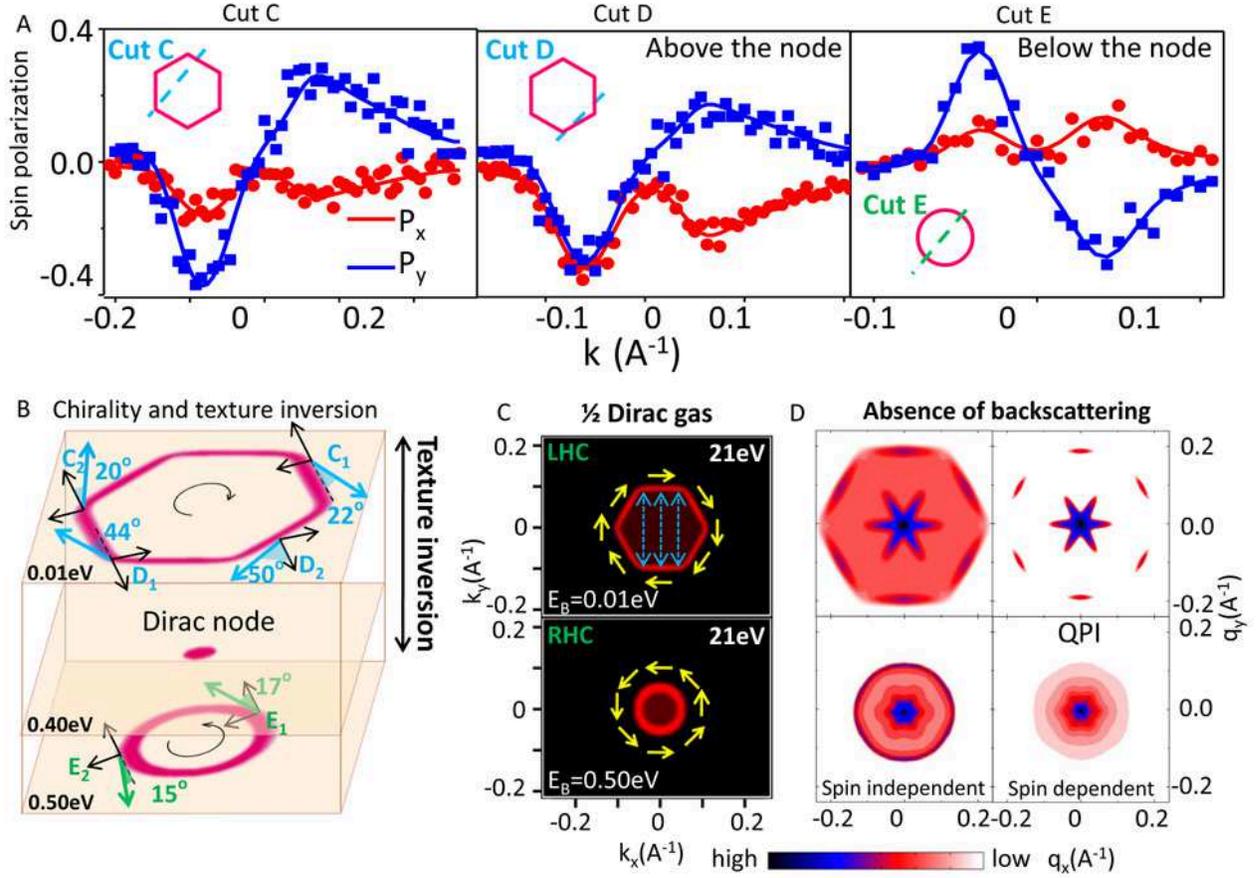}
\caption{\textbf{Chirality inversion and spin dependent scattering profiles for charge and spin transport.} \textbf{(A)} Measured in-plane spin-polarization profiles of cuts C, D and E (Fig. 2). \textbf{(B)} Fitted in-plane azimuthal angle values of spin vectors measured along cuts C, D and E on the ARPES measured FSs. Spin rotation handedness or chirality changes from lefthanded to right-handed in passing through the Dirac node toward the higher binding energy. The binding energies are indicated at the bottom-left corner for the experimental FSs presented. \textbf{(C)} ARPES measured FSs are shown with spin directions based on polarization measurements. L(R)HC stands for left(right)-handed chirality. Photon energy used for spin-resolved measurements is indicated at the top-right corners. \textbf{(D)} Spin independent and spin dependent scattering profiles on FSs in \textbf{(C)} relevant for surface quasi-particle transport are shown which is sampled by the quasi-particle interference (QPI) modes.}
\end{figure*}
\begin{figure*}
%\centering
\includegraphics[width=16cm]{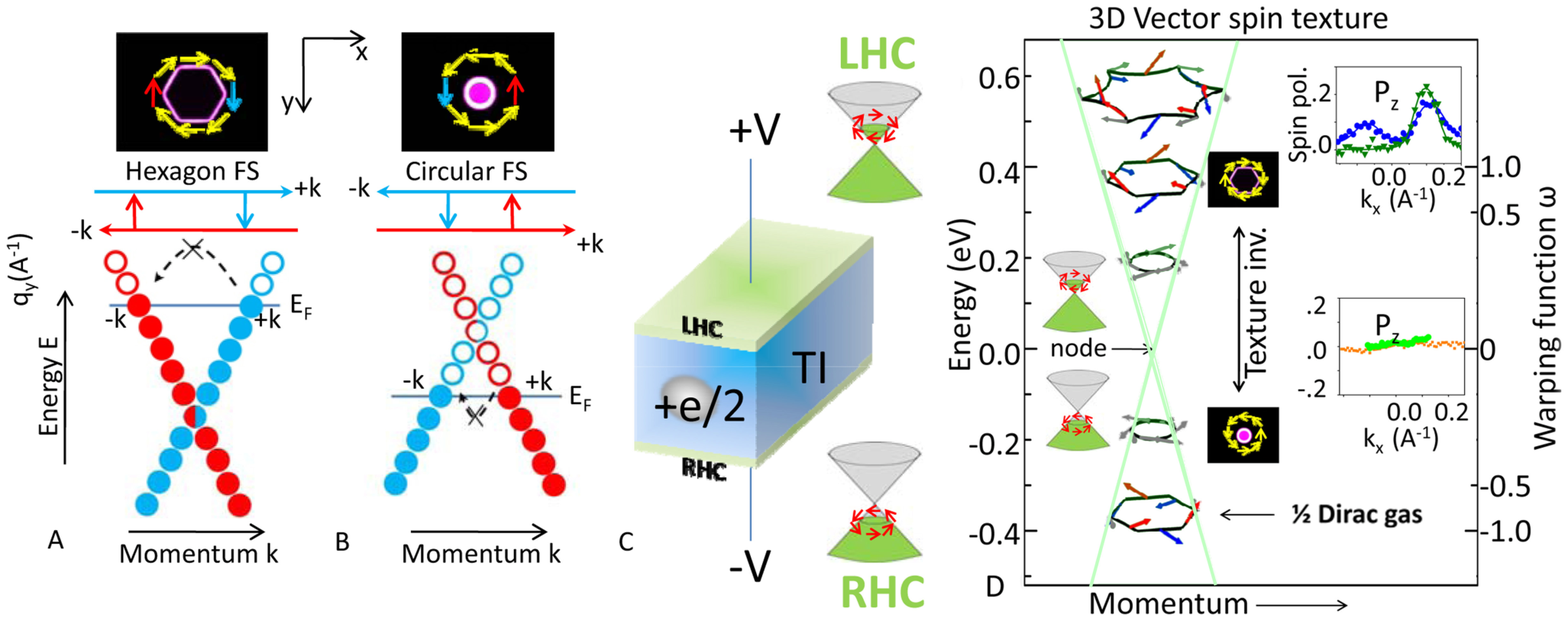}
\caption{\textbf{Inversion of spin-momentum locking profile and spin texture phase diagram.} \textbf{(A} and \textbf{B)} A $180^{\circ}$ turn-around of the spin-momentum locking profile is observed when chemical potential moves across the Dirac node. In the present experiment, this is equivalent to measuring the profile above and below the Dirac node. The profiles above and below the Dirac node are shown in \textbf{(A)} and \textbf{(B)} respectively. \textbf{(C)} A
capacitor-like device geometry using topological insulator thin films of opposite chirality (LHC and RHC) composite is predicted to host exotic exciton properties where quasi-particle excitation carries fractional charge. \textbf{(D)} Spin texture evolution of topological surface bands as a function of energy away from the Dirac node (left axis) and geometrical warping factor $\omega$ (right axis). The warping factor is defined as $|\omega|={\frac{k_F(\bar{\Gamma}-\bar{M})-k_F(\bar{\Gamma}-\bar{K})}{k_F(\bar{\Gamma}-\bar{M})+k_F(\bar{\Gamma}-\bar{K})}} {\times} {\frac{2+\sqrt{3}}{2-\sqrt{3}}}$, where $\omega=0$, $\omega=1$, and $\omega>1$ implies circular, hexagonal and snowflake-shaped FSs respectively. The sign of $\omega$ indicates texture chirality for LHC (+) or RHC (-). Insets: out-of-plane 3D spin-polarization measurements at corresponding FSs.}
\end{figure*}
\end{document}